\documentclass{aa}
\usepackage{graphicx}
\begin{document}
\title{Motivation and possibilities of affordable low-frequency radio interferometry in space}

\subtitle{Applications to exoplanet research and two instrument concepts}

\author{P. Janhunen \inst{1} \and A. Olsson\inst{2} \and R. Karlsson\inst{2}
\and J.-M. Grie{\ss}meier \inst{3}}

\offprints{P. Janhunen}

\institute{Finnish Meteorological Institute, Geophysical Research (FMI/GEO), 
           POB 503, FIN-00101 Helsinki, Finland\\
           \email{pekka.janhunen@fmi.fi}
       \and
           Swedish Institute of Space Physics, Uppsala Division,
           POB 537, SE-75121, Uppsala Sweden\\
           \email{ao@irfu.se}
       \and
           Institut f\"ur theoretische Physik,
           TU Braunschweig,
           Mendelssohnstrasse 3, D-38106 Braunschweig, Germany\\
           \email{j-m.griessmeier@tu-bs.de}
}

\date{Received April, 2003; accepted ????, 200?}

\abstract{
The motivation to build spaceborne interferometric arrays for
low-frequency radio astronomy is widely recognised because frequencies
below the ionospheric cutoff are inaccessible for ground-based radio
telescopes.  We discuss the theoretical possibilities to use
low-frequency spacecraft arrays to detect signals from magnetized
extrasolar planets, including earthlike ones. A major uncertainty
that prohibits us from knowing if it is possible to detect exoplanet
cyclotron maser signals is the incomplete knowledge of the properties
of the interstellar plasma.

We also present some ideas of how to construct efficient and
affordable space-based radio telescopes. We discuss two possibilities, a
log-periodic antenna in the spin plane and a two-spacecraft concept
where one spacecraft holds a large parabolic wire mesh reflector and
the other one contains the receiver. In the latter case, the effective
area could be of the order of 1 km$^2$. The purpose of the paper is to
stress once more the importance of spaceborne low-frequency
measurements by bringing in the intriguing possibility of detecting
earthlike exoplanet radio emissions and to demonstrate that building
even very large low-frequency antennas in space is not necessarily too
expensive.
}

\maketitle
%


\section{Introduction}

Because of the existence of ionospheric plasma, radio frequencies
below 10-30 MHz are difficult or impossible to study using
ground-based radio telescopes, whereas low-frequency space-based
interferometric arrays with good angular resolution have not yet been
built. There are many interesting phenomena to study by low-frequency
radio astronomy. This has been pointed out in many references, e.g.,
Basart \cite{BasartEtAl97a}b and Kassim and Weiler
\cite{KassimAndWeiler90}.  These references also give comprehensive
lists of radioastronomy research topics for low-frequency space
interferometry, which will not be repeated here. Additionally, activities
have started lately to make the first radio measurements of extrasolar
planets (exoplanets). A more detailed investigation of this is a subject
of this paper (section 2 below). Before going into the details of radio detection of
exoplanets, we list briefly the principal possibilities to study
exoplanets in general (Clark \cite{Clark98}, for recent short review, see Schneider \cite{Schneider2002}):

\begin{enumerate}

\item Detecting the planet-induced wobbling motion of the parent star
from the Doppler effect it induces in its spectral lines (Mayor and
Queloz \cite{MayorAndQueloz95}). The method gives estimates for the
mass and orbiting distance of the planet. This is the primary method
for detecting new exoplanets at the moment.

\item Detecting the wobbling motion astrometrically (Benedict et al. \cite{BenedictEtAl2002}).

\item Observing transit situations where the planet moves in front of
the parent star and produces a small change in the light curve of the
star (Henry et al. \cite{HenryEtAl2000}, Charbonneau et
al. \cite{CharbonneauEtAl2000}). One can find the orbiting distance
and the planetary radius by this method, and in principle also obtain
information about the atmospheric composition of the planet. The
method requires that the orbital plane is favourably aligned, so it
cannot be used for systematic searches in the local neighbourhood.

\item Using gravitational microlensing events caused by remote
exoplanets (Bennett and Rhie, \cite{BennettAndRhie2000}). This
technique has the potential of providing statistics of exoplanet
orbital distances and masses in the galaxy.

\item Direct observation of a nearby exoplanet in visual or infrared
(the DARWIN mission, {\tt http://sci.esa.int/home/darwin/index.cfm},
Fridlund and Kaltenegger \cite{FridlundAndKaltenegger2002}). Like the
transit technique, this method gives the size and orbital distance,
and in its more sophisticated forms also information about the
composition of the exoplanet atmosphere. It allows for a systematic
study of all exoplanets in the local neighbourhood.

\item Optical or infrared observation of an exoplanet with the help of
diffraction at the dark limb of the moon (Richichi \cite{Richichi2003}).

\item Direct observation of the exoplanet's radio signal. This will
give the magnetic field of the planet. Like the visible and infrared
direct observation, the method is suitable mainly for nearby exoplanets.

\end{enumerate}

The last method listed is the one elaborated on this paper. It is the
only method capable of giving information about the exoplanet's
magnetic field and it has not been as extensively developed as some
other methods. When combined with direct infrared or optical methods
it allows us to study a given exoplanet more closely than with the
other methods mentioned. In section 2 we discuss the theoretical
possibilities for exoplanet radio detection and in section 3 we
discuss two possible affordable but powerful antenna configurations in
space. We close the paper with a summary.

\section{Exoplanet cyclotron maser emissions}

In the radio frequencies there have been a few attempts to detect the
cyclotron maser emissions from giant gas exoplanets (Winglee et
al. \cite{WingleeEtAl86}, Bastian et
al. \cite{BastianEtAl2000}, Zarka et al. \cite{ZarkaEtAl97}), although
no exoplanet radio signal has yet been detected.
These searches were ground-based and thus they were
limited to frequencies above the ionospheric cutoff. Actually, the frequencies
used thus far have been quite high: for example the lowest frequency
used by Bastian et al. (\cite{BastianEtAl2000}) was 74 MHz (the highest
frequency at Jupiter is 39.5 MHz). The same frequency was probed with
0.12 Jy sensitivity again in 2002 with no exoplanet detected (Farrell
et al. \cite{FarrellEtAl2003}).

In Earth's case, the cyclotron maser emission is called the auroral
kilometric radiation (AKR) and its frequency range is 50-800 kHz
(Gurnett \cite{Gurnett74}), i.e. much lower than the ionospheric
cutoff of $\sim 10$ MHz. The total AKR power emitted by the Earth is $10^7-10^8$ W on the
average (Gallagher and Gurnett \cite{GallagherAndGurnett79}) and $10^9$ W during intense
magnetic storms (Gurnett \cite{Gurnett74}, \cite{Gurnett91}), which means that AKR
power is of the order of one percent of the particle precipitation
power.  The peak AKR power is therefore about $2\times 10^8$ times
less than the total solar radiation power incident on the top of the
atmosphere ($1.7\times 10^{17}$ W). Due to the radiation balance, the
latter number, $1.7\times 10^{17}$ W, is equal to the total power that
the Earth radiates away in infrared; thus the maximum outgoing AKR
power is $\sim 2\times 10^8$ times smaller than the outgoing visible
and infrared radiation. As the energy of a photon is inversely
proportional to its wavelength, the energy of an AKR photon (typically
3 km, corresponding to 100 kHz) is $3\times 10^8$ times smaller than
the energy of a typical infrared photon emitted (10 $\mu$m,
corresponding to 300 K). Since the number of radiated photons per time
is equal to the output power divided by the photon's energy, the
maximum number of AKR photons is likely to be comparable to the number
of photons emitted in the infrared. The infrared emission has been
proposed to be used for possible detection of earthlike extra-solar
system planets (exoplanets) using interferometric techniques.
As an example, if an earthlike exoplanet is at 10 parsec distance and
it radiates 100 kHz photons with 1 GW power, the AKR photon flux
is $1.5 \times 10^{37}$ s$^{-1}$. In one second, $\sim 10^7$ of these
photons pass through a one square kilometer area near the Earth.

The photon count estimate thus shows that in principle, receiving
auroral radio emissions from an earthlike exoplanet should be possible
and not more difficult than receiving its infrared signal. If a radio
signal from an earthlike exoplanet would be received, it might
significantly increase our knowledge of such planets in general,
because the frequency of the emission is the electron gyrofrequency
($f/{\rm Hz} = 28 B/{\rm nT}$) and we can thus get knowledge of the
magnetic field at the bottom of the planet's auroral acceleration
region, thus giving a lower limit for the surface magnetic field (Wu
and Lee \cite{WuAndLee79}). How quickly the emission varies in time
reflects the time constant of the substorm cycle (substorms occur at
unpredictable and quasi-random times, but the number of substorms over
a long enough observation period should depend only on the avearge
properties of the solar wind and the size of the magnetosphere), which is related to
the size of the magnetosphere. At least if the exoplanet is
sufficiently earthlike, it should have an identifiable substorm
cycle. If a planet has a significant internal magnetic field, it must
be produced by a dynamo working in a liquid core. The dynamo action
needs a Coriolis force to work, thus the presence of a magnetic field
tells that the planet is rotating fast and that it has a liquid
core. If the planet is otherwise earthlike and with suitable surface
temperature (Franck et al. \cite{FranckEtAl2000}), these conditions
increase the likelihood that it could have conditions suitable for
life. Fast enough rotation keeps the day/night temperature difference
within reasonable limits and the presence of a liquid core is
necessary, although not sufficient, to have plate tectonics to work
(Bostrom \cite{Bostrom2000}). Plate tectonics is, on the other hand,
probably required to recycle carbon back into the mantle and thus
prevent a runaway CO$_2$ greenhouse effect that would turn the planet
inhabitable (Gonzales et al. \cite{GonzalesEtAl2001}, Franck et al., \cite{FranckEtAl99}, \cite{FranckEtAl2000}). Thus, measuring the radio emissions could be a way to
investigate if a planet which is known to be roughly earthlike from
other measurements is indeed habitable, or even to make such a
judgement without being able to detect the planet at all by other
means.

There are two principal difficulties in detecting exoplanet radio
signals:

\begin{enumerate}

\item Interstellar and interplanetary plasma distort the radio waves,
limiting the maximum angular resolution that can be reached (Linfield
\cite{Linfield96}). If the angular resolution is not good enough, the
exoplanet's radio signal is lost in the diffuse background emission,
independent of how large radio telescope is used. The amount of
diffuse background emission is, however, essentially unknown: what is
known is a very low angular resolution map (Brown \cite{Brown73}) and
it is not known to what extent the map comes from truly diffuse
emissions and to what extent localised sources. If it comes from
localised sources, the possibilities for detecting exoplanets increase
correspondingly. The only way to find it out is to build a spaceborne
low-frequency interferometer.

\item The maximum angular resolution set by the interstellar plasma is
in any case almost certainly worse than what would be required to
separate the planet from its parent star. Thus in practice the whole
stellar system will be seen as one point source. If the parent star or
the giant planets outshine the earthlike exoplanet in the radio
frequency used, detecting the planet's own signal may then be
difficult. However, during certain times even the Earth's AKR may
outshine the radio emissions of the Sun so this is not necessarily a
problem. For cyclotron emissions of Jupiter-like exoplanets the
problem is less severe anyway. It means, however, that attention must
be given to methods to distinguish planetary and stellar radio
emissions. One way to help distinguishing them is that planetary emissions are
circularly polarised while the stellar emissions are not.

\end{enumerate}

To make progress towards solving both problem areas, new high
sensitivity and high angular resolution measurements are needed.
These are only possible to obtain using spaceborne low-frequency
interferometers.


Even if these studies should show that it is not possible to detect
earthlike exoplanets due to one of the problems listed above, these
new instruments would greatly contribute to the ongoing search for
radio signals from gaseous exoplanets. The reason is twofold. Firstly,
a much better sensitivity can be expected for a spaceborne
interferometer than for ground-based telescopes. The sensitivity
mainly depends on the baseline, which can be selected according to the
requirements (see also section 3.4.). The radio flux density at the Earth
is estimated by Farrell et al. \cite{FarrellEtAl99}.  Secondly, the
ionospheric cutoff can be avoided. The maximum frequency emitted by a
planet is equal to the gyrofrequency on the planet's surface. Planets
orbiting their host stars in very close orbits (i.e. $d\le 0.1$ A.U.)
are subject to strong tidal dissipation, leading to gravitational
locking. For such planets the rotation period equals the orbital
period, and fast rotation is not possible. Commonly employed
scaling-laws for the planetary magnetic moment (see,
e.g. \cite{CainEtAl95}) which are usually based on an
$\alpha\omega$-dynamo always yield a magnetic moment rapidly
decreasing with increasing rotation period. This puts the expected
emission frequencies of several planets (51 Peg, $\nu$ And, 55 Cnc)
well below ionospheric cutoff \cite{FarrellEtAl99}. Furthermore, as the
emitted flux density decreases with increasing frequency, even those
planets which do emit at higher frequencies will be easier to detect
at lower frequencies \cite{Rucker2002}.

\section{Low-frequency antenna design in space}

A lot of work has been done to design a low-frequency array in space
under the ALFA project (Jones et al. \cite{JonesEtAl2000}). While this
body of work provides useful background information, we shall take a
fresh approach which takes into account exoplanet detection and recent
developments in space technology. This is the reason why the
technical solutions we will come up with further below are different
from those proposed in the ALFA project.

To reach good angular resolution at low frequencies, interferometry
with multiple spacecraft is needed, while the number of spacecraft
should be kept at minimum to make the mission less expensive. To increase
the ability of the system to observe faint sources, either the number of
interferometric measurement points (spacecraft) can be increased or
the antenna on each spacecraft can be made more directing. In this
paper we especially concentrate on the latter option because it
appears more promising in terms of cost-effectiveness. We
shall also assume that a rather wide frequency band (e.g., 0.3-30 MHz)
must be covered; this precludes designing the antennas specifically
for one frequency.

In order to be practical and affordable, at least the following
technical requirements must be met. The total mass of the antenna must
not exceed a few hundred kilograms. The antenna must be such that it
can be packed densely during launch and deployed in space. The construction must
not be overly sensitive to micrometeor damage. The length of stiff
structures must be kept below a few tens of meters.

Based on these requirements, we considered and rejected many
design ideas. A simple dipole wire antenna is technically easy
to deploy and is very lightweight (the mass of one kilometre of wire
can be made as small as $\sim 1$ kg). For wavelengths larger than the
dipole wire length, a dipole antenna works well. For shorter wavelengths the gain pattern
becomes complicated and the effective area is of the order of
wavelength squared. Thus by using a set of independent dipoles of
different lengths (lengths e.g. scaling like powers of two), a wanted
frequency range can be covered. This may not be too bad
in terms of effective area, but a drawback of dipoles is that they are
hardly directing at all. In theory, directing antennas are not needed
since a similar effect can be produced digitally if the number of
measurement points is large enough, but in practice the finite dynamic
range of A/D converters makes it impossible to detect very faint
source in the presence of strong radio ``noise'' from Earth's AKR,
Jupiter and the Sun unless the antenna elements themselves are also
directing.

The surface of a large gas-filled balloon could be used as an antenna
or a spherical reflector, but such a construction is vulnerable to
micrometeors (the gas would escape through the holes made by the
meteors). A spinning disk-shaped thin mylar or kapton membrane
(dielectric) with conducting antenna patterns wired on it is too heavy
even if made from 1 $\mu$m membrane. Frequency-independent
spiral-shaped wire antennas could be periodically realised in the spin
plane by adjusting the spin rate of the spacecraft periodically, but
their directivity and effective area is not so much better than those
of a simple dipole wire antenna that the extra technical effort
involved would be warranted. A Yagi-Uda antenna can be made very
directing, but its frequency range is too narrow to be useful.

\subsection{Planar log-periodic spin-plane antenna}

\begin{figure}
\centering
\fbox{\includegraphics[width=8.4cm,bb=50 200 550 550,clip=true]{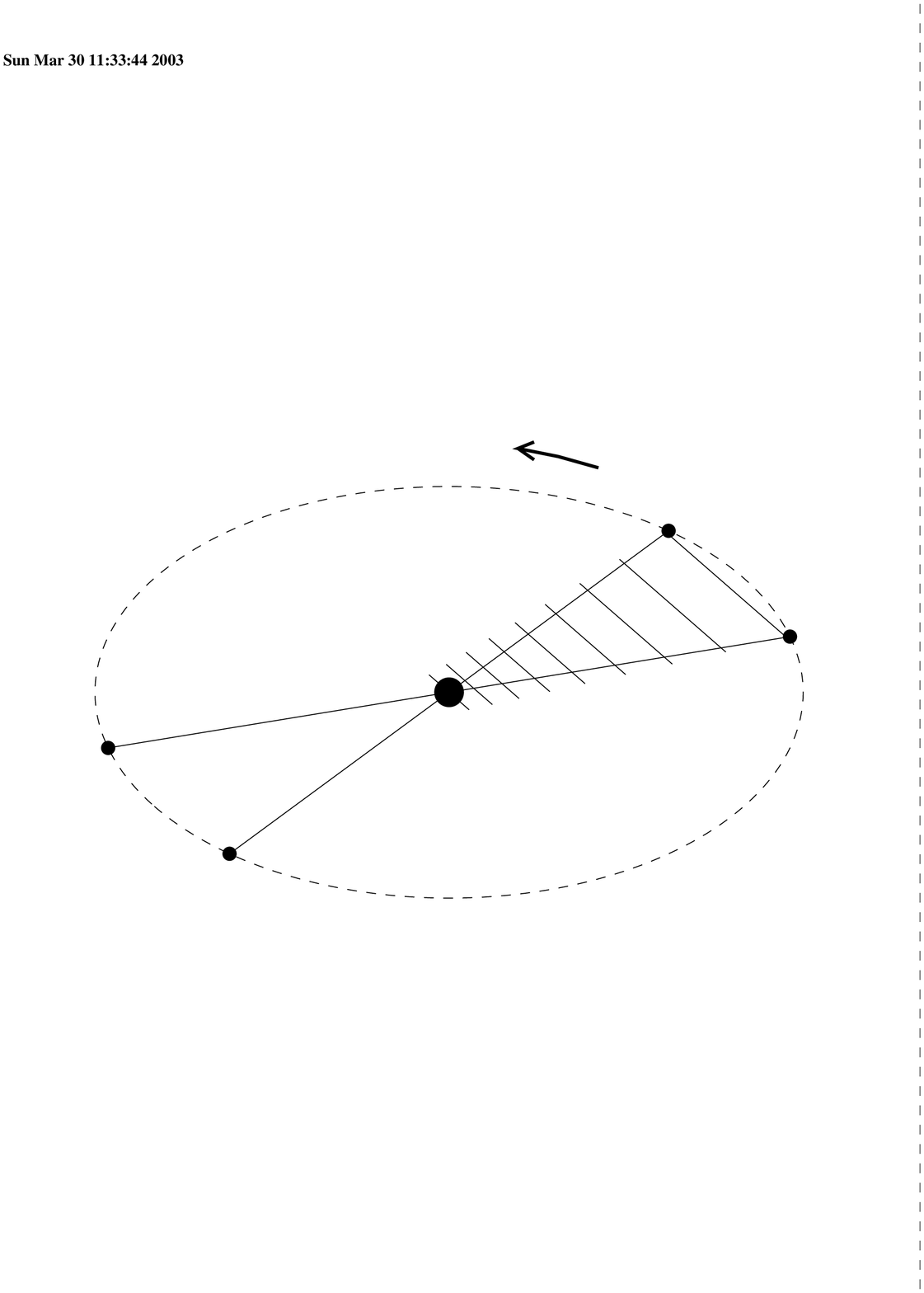}}
\caption{Log-periodic antenna in the spin plane
}
\label{fig:logperiodic}
\end{figure}

Figure \ref{fig:logperiodic} shows a log-periodic wire antenna in the
spin plane, spanned by rotating masses. The antenna itself is only on
one side of the spacecraft, but for mechanical stability, masses are
added on the other side as well. The radius of the dashed circle
should be $\sim 2$ km (the length of the outermost antenna wire should
be one half the longest wavelength wanted). Deploying such a system is
more complicated than deploying ordinary dipole antennas, but should
not be too difficult. The thin antenna wires are stretched between the
radial support wires which are kept in shape by the rotating
masses. For optimal operation the antenna wires should have a definite
thickness which depends on their length (Balanis
\cite{Balanis97}). Using such thick wires is out of question, but a
similar effect could probably be obtained without increasing the
antenna mass by using thin conducting bands. A band is also expected
to be more micrometeor-resistant than a wire since a meteor only
punches a hole in it but does not break the band like it breaks the
wire in case of a hit. Thus the meteor danger should not prohibit the
use of kilometer-long wires. The feasibility of deploying
and maintaining long wires has also been demonstrated in practice. A 20 km long tether has
been almost sucessfully deployed from the Space Shuttle in 1996,
although the tether broke due to ohmic heating produced by an induced
current in the near-Earth high magnetic field (high induced currents
occur only at low Earth orbit where the magnetic field is
large). Furthermore, the IMAGE satellite has flown with 500 m
tip-to-tip booms for two years now (Burch \cite{Burch2003}).

A log-periodic antenna operates well over a wide wavelength range
determined by the minimum and maximum length of the antenna rods.  The
directivity of a log-periodic antenna is usually 7-10 dB (Balanis
\cite{Balanis97}). For a maximum wavelength of 600 m (500 kHz) the antenna
length should be $\sim$ 2 km (the exact length depends on the amount
of directivity and on the frequency range), having a total amount of
wire of $\sim 10$ km. A 0.4 mm diameter wire with aluminium mass
density (2700 kg m$^{-3}$) has mass per length of 0.34 kg/km so the
total mass of wires would be a few kilograms. Even together with the
rotating masses that keep the support wires stretched, the antenna mass
would remain well below 100 kg.

The object to be measured with a spinning log-periodic antenna should
be in the spin plane and useful data of that object can be taken only
a short period during each spin. The effective area of a log-periodic
antenna is similar to an optimal dipole at each frequency so the chief
benefit of the log-periodic construction is its directivity.

\subsection{Parabolic ``spidernet'' reflector with two spacecraft}

\begin{figure}
\centering
\fbox{\includegraphics[width=8.4cm,bb=0 608 280 830,clip=true]{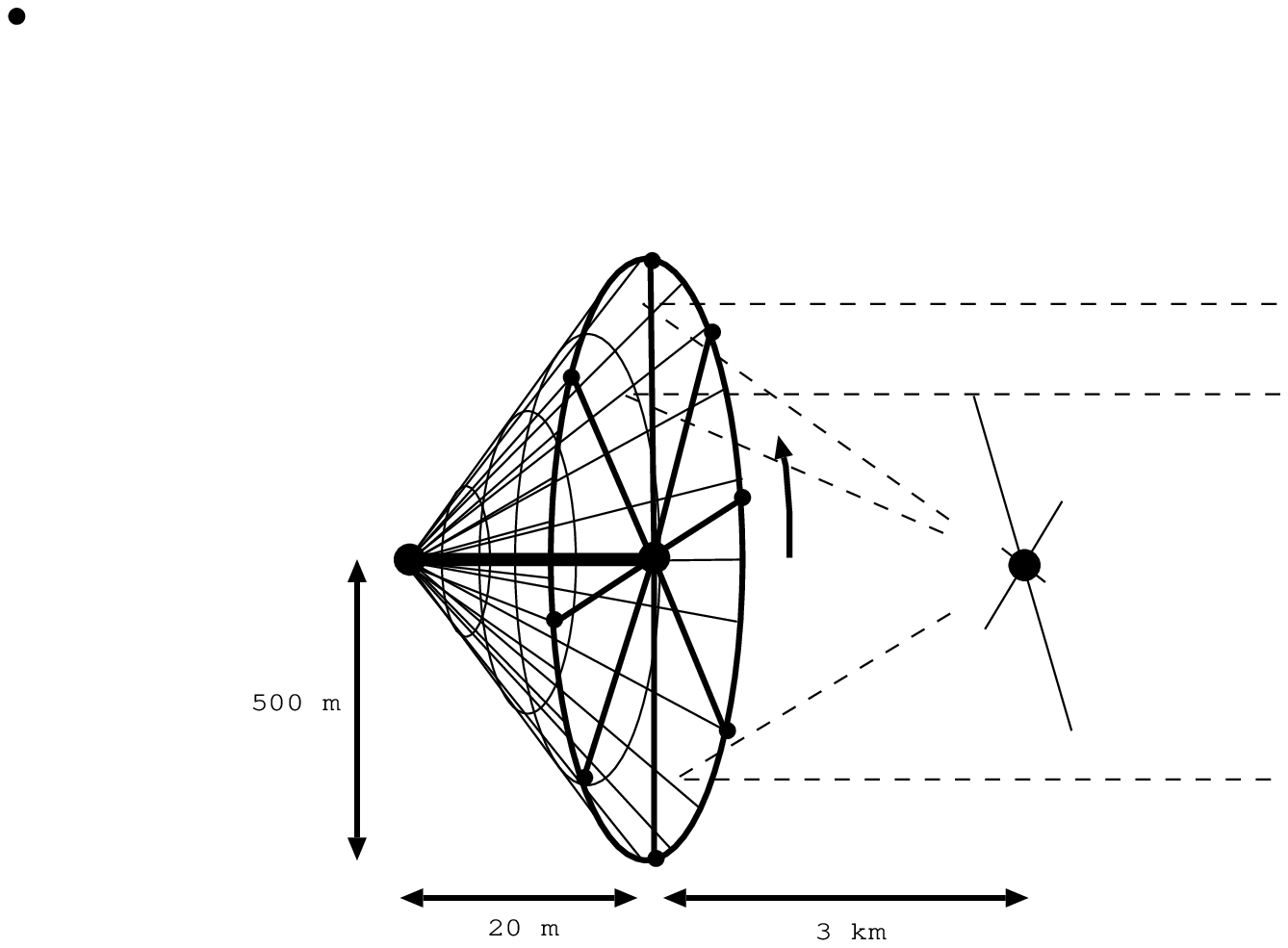}}
\caption{Two-spacecraft ``spidernet'' reflector and receiver in space.
Here the reflector is shown as conical for simplicity. Support threads
and drawn with thicker line than the reflecting spidernet wires.
The figure is not to scale.
}
\label{fig:spider}
\end{figure}

It is not possible to increase the effective area of an antenna
element much beyond the square of the wavelength without using a
collector, which must be some type of a parabolic reflector in
practice. A parabolic reflector has a very good directivity and an
effective area which is roughly equal to its physical area. It works
for wavelengths shorter (preferably: much shorter) than its radius.

A circular disk made of wires (a ``spidernet'', Fig.~\ref{fig:spider})
can be kept in shape by spinning, and it acts as a reflector for waves
whose wavelength is longer than the wire mesh spacing. To collect the
radiation, however, the dish must be bent, which necessitates a long
(half the dish radius) solid axis. For example, if the dish radius is
500 m, the solid axis should be 250 m long. Constructing such a solid
axis and deploying it in space is a technological challenge, although
probably not impossible since the force that the axis has to withstand
is not large.

The length of the required solid axis required to support the
parabolic dish decreases if the dish is made more planar, but then the
focal point moves away from the spacecraft. Using two spacecraft,
one holding the reflector dish and the other hosting the receiving
antenna looks more promising, however. For example, with a 20 m long solid
spinning axis one can keep in shape a parabolic dish of 1 km
diameter. The focal point (the receiver spacecraft) is then 3 km away.

Assuming a spidernet wire mesh with $d=5$ m wire spacing, frequencies
up to $\sim 50$ MHz are reflected. If the spidernet radius is $R=500$
m, the area is $A = 8 \times 10^5$ m$^2$ and the total length of wire
required to weave the net is $L = 2A/d = 300$ km. If the wire mass per
length is the same as given above (0.34 kg/km), the total wire mass
is then 100 kg. An equivalent solid membrane dish of the same area
would be much heavier, or about 2000 kg if one uses a very thin
membrane of 1 $\mu$m thick and with aluminium mass density. The mass
estimate of 100 kg was based on wire diameter 0.4 mm. Probably most of
the wires could be thinner because they are repeated every 5 m; a few
tens of kilograms might be a realistic estimate for the spidernet mass. On the other hand, to
keep the spidernet in its shape it needs to be rotated and a mass
which is a few times larger than the spidernet mass must reside on the
circumference. The actual mass of the reflector construction depends
on the adopted spin rate, the wanted upper frequency limit and the
employed materials, but the above consideration shows that reaching a
mass of as small as 100 kg is not out of the question.

The shape of the parabolic spidernet reflector needs not be very
accurate at the long wavelengths and long focal length considered
here. A numerical experimentation using the NEC code shows that even a
cone-shaped reflector works almost as well as a parabola for the
parametre values discussed above. As is well-known for parabolic
dishes, the directivity of the antenna as a whole is low for low
frequencies and increases up to the cutoff frequency which is
determined by the mesh spacing of the spidernet. For example, with
dish radius of 500 m, a directivity of 16 dB is obtained at 4 MHz.

The deployment of a spidernet with its support wires
(Fig.~\ref{fig:spider}) should occur automatically by the centrifugal
force. The technical construction of the reflector spacecraft is made
easier by the fact that it does not need many other devices or
outstanding instruments.

\subsection{Receiver cooling}

Based on currently existing measurements of the galactic radio noise
at low frequencies, cooling of the receiver should be unnecessary
since the sky temperature is much larger than 300 K. However, since
accurate maps of the radio sky at low frequencies have not been made,
including a simple passive radiator cooling down to $\sim 30$ K might
still prove valuable since the background emission may turn out to
consist of pointlike sources in some regions of the sky.

\subsection{Orbit selection and baseline for interferometry}

To avoid manmade signals, a radioastronomical spacecraft should be
either orbiting the Moon so that it is periodically behind it or it
should be as far away from the Earth as possible, possibly even
orbiting the Sun. In the Moon orbit case, data must be stored in the
spacecraft and transmitted to Earth when the eclipse period stops, or
it must be relayed to Earth by yet another spacecraft. Nowadays it is
no problem to have sufficient memory onboard to store eclipse period
data, so the Moon orbit is attractive because of its other benefits,
e.g. cheaper communication link to Earth because of the relatively
short distance. However, if the wanted interferometer baseline is
longer than the radius of the Moon, the Moon orbit cannot be used. The
maximum feasible baseline is determined by the amount of scattering
and bending caused by interstellar and interplanetary plasma. The
properties of interstellar plasma (density and density gradients) are
still largely based on models rather than data (Rickett
\cite{Rickett90}, Linfield \cite{Linfield96}), so accurate answers do
not exist. It has been estimated by the ALFA proposal scientists that
baselines longer than 100 km are not useful in the 0.03-30 MHz range (Jones et
al. \cite{JonesEtAl2000}); however, this estimate is based on the
assumed average properties of the interstellar plasma in the
kiloparsec scale. It may be that especially in our local neighbourhood
($<30$ parsec) there are at least some directions where the interstellar
plasma is tenuous and smooth enough to allow for higher angular
resolution than the 10 arcmin what is provided by a 100 km baseline at
1 MHz. Especially when trying to observe exoplanets, the local
neighbourhood is important, and there an angular
resolution as high as possible is required, so it is our suggestion that the
baseline should preferably be 500--1000 km, giving angular resolution
of 1-2 arcmin at 1 MHz. Since the plasma effects decrease as $1/f^2$
and the angular resolution as $1/f$, at
high frequencies the bottleneck is the spacecraft
separation rather than plasma effects.

\section{Summary and conclusions}

The purpose of this paper was to continue and refresh the interest
towards low-frequency radio astronomy by pointing out one intriguing
possible application area (exoplanets) and by outlining two possible
technical constructions (one simple and one more ambitious) of a
single-satellite antenna with high directivity. Detecting nearby
earthlike exoplanets with radio methods is certainly very difficult,
but the same is true for any other method thus far proposed. Because
the limits of the radio method depend strongly on the properties of
the interstellar plasma, which are currently not accurately known, the
way forward is to try and implement the method. In the best case one
could study the same exoplanets by both optical, infrared and radio
methods; while each method used alone is incomplete, they are much
more powerful when used together.

Regarding the technical methods to construct spaceborne low-frequency
antennas, the simpler solution, a log-periodic antenna in the spin plane is
lightweight and relatively easy to construct. Its main drawback is a
modest effective area and the fact that the instrument cannot be
continuously pointed towards an object (this could also be a benefit if
several objects are to be scanned during the spin).

The more ambitious solution, but still quite affordable and feasible
using current technology, is a large spidernet parabolic network
paired with a separate receiver satellite containing e.g. dipole
antennas. To reflect low-frequency waves, a coarse mesh of thin
conducting wires is enough, which makes it possible to construct very
large dish reflectors with radius 500-1000 m with a mass as small as 100
kg.

Since the purpose of the paper is not to present a concrete new
mission, there are many technical details that we did not investigate,
for example, the optimal antenna configuration (set of
dipoles, or perhaps spiral antennas) in the receiver satellite of the
spidernet configuration. By bringing in the exoplanet viewpoint we
wanted to exemplify that low-frequency arrays in space would be
interesting also for magnetospheric and auroral physicists outside the
traditional domain of radioastronomy. Interpreting an exoplanet
radio data in an optimal way
requires a detailed knowledge of magnetospheric physics and the
ability to quantitatively simulate magnetospheric and auroral
processes in conditions that are not exactly similar to any of the
examples we have studied in the solar system.

\begin{acknowledgements}
We are grateful to Lennart {\AA}hl\'en, Bo Thid\'e, Kaj
Wiik and Esko Valtaoja for useful information and discussions.
\end{acknowledgements}

\end{document}